\documentclass{sigchi}

\usepackage{balance}  
\usepackage{graphics} 
\usepackage{times}    
\usepackage{url}      

\makeatletter
\def\url@leostyle{%
  \@ifundefined{selectfont}{\def\UrlFont{\sf}}{\def\UrlFont{\small\bf\ttfamily}}}
\makeatother
\def\pprw{8.5in}
\def\pprh{11in}

\usepackage[pdftex]{hyperref}
\hypersetup{
pdftitle={SIGCHI Conference Proceedings Format},
pdfauthor={LaTeX},
pdfkeywords={SIGCHI, proceedings, archival format},
bookmarksnumbered,
pdfstartview={FitH},
colorlinks,
citecolor=black,
filecolor=black,
linkcolor=black,
urlcolor=black,
breaklinks=true,
}

\begin{document}

\title{Enhancing Visual Fashion Recommendations with Users in the Loop}
\numberofauthors{5}

\numberofauthors{5} 
\author{
Anurag Bhardwaj, Vignesh Jagadeesh, Wei Di, Robinson Piramuthu, Elizabeth Churchill\\
       \affaddr{eBay Research Labs}\\
       \affaddr{2145 Hamilton Ave., San Jose, CA, USA}\\
       \email{{anbhardwaj,vjagadeesh, wedi, rpiramuthu, echurchill}@ebay.com}       
}

\maketitle

\begin{abstract}
We describe a completely automated large scale visual
recommendation system for fashion. Existing approaches have primarily relied on purely computational models to solving this problem that ignore the role of users in the system. In this paper, we propose to overcome this limitation by incorporating a user-centric design of visual fashion recommendations. Specifically, we propose a technique that augments ``user preferences" in models by exploiting elasticity in fashion choices. We further design a user study on these choices and gather results from the ``wisdom of crowd" for deeper analysis. Our key insights learnt through these results suggest that fashion preferences when constrained to a particular class, contain important behavioral signals that are often ignored in recommendation design. Further, presence of such classes also reflect strong correlations to visual perception which can be utilized to provide aesthetically pleasing user experiences. Finally, we illustrate that user approval of visual fashion recommendations can be substantially improved by carefully incorporating these user-centric feedback into the system framework.
\end{abstract}



\section{Introduction}

Fashion is widely considered as a mirror into social, cultural and economical status of a society. Often interpreted as a ``non-verbal communication" mode with the society~\cite{Lurie:1981:TLC}, it is also one of the most studied problems of our times. Numerous attempts have been made at understanding intricacies of fashion by sociologists, cognitive psychologists as well as economists~\cite{Cholachatpinyo:2002:FMM,Moody:2010:FMM}. These studies have led to a better understanding of why certain fashion choices (color palette, pattern, fabric design) have dominated at some points in time and how some of these choices may have been choreographed by forces outside of a typical user's control (i.e. color tone and pattern design set by a cartel of designers). 

Recently, new techniques have emerged in computer science and vision science in particular, that have provided tools to utilize massive amounts of image data on web~\cite{YamaguchiCVPR12}. Using such techniques, it is now possible to study these choices from a user's perspective and get a sense of what people like ``on the ground". This allows us to take a bottom up approach to understanding Fashion, both personally (what YOU like) and collectively (what WE like) that can be highly useful for a number of applications. In this paper, we propose to study these approaches in context of ``Fashion Recommendation" also commonly known as ``Fashion Co-ordination". A fashion recommender is usually defined as an automated system that finds all co-ordinating pieces of an outfit given a particular piece as an input. For instance, given a blue skirt as an input, a fashion recommender would provide all tops that co-ordinate well with the given skirt. This problem has variety of applications in commerce one of which is ``up-selling" where an online shopper with an fashion item in her shopping cart is recommended other fashion items that co-ordinate the cart. This application is also appealing from a user perspective since a key issue in commerce platforms with large inventories is to be able to support visual search with filters that include pattern taste for different scenarios such as \textit{direct match} (i.e. leopard to leopard), \textit{for theme} (i.e. leopard to leaves for jungle theme) and \textit{for complement/co-ordinate} (i.e. leopard to block color to accent the leopard). In other words, where word descriptors can fail, visual matches can succeed. For instance, polka dots by size are hard for humans to describe in words but perceptually easy to recognize.\\
\begin{figure}[t]
\centering
\begin{tabular}{c}
\includegraphics[width=1.0\linewidth]{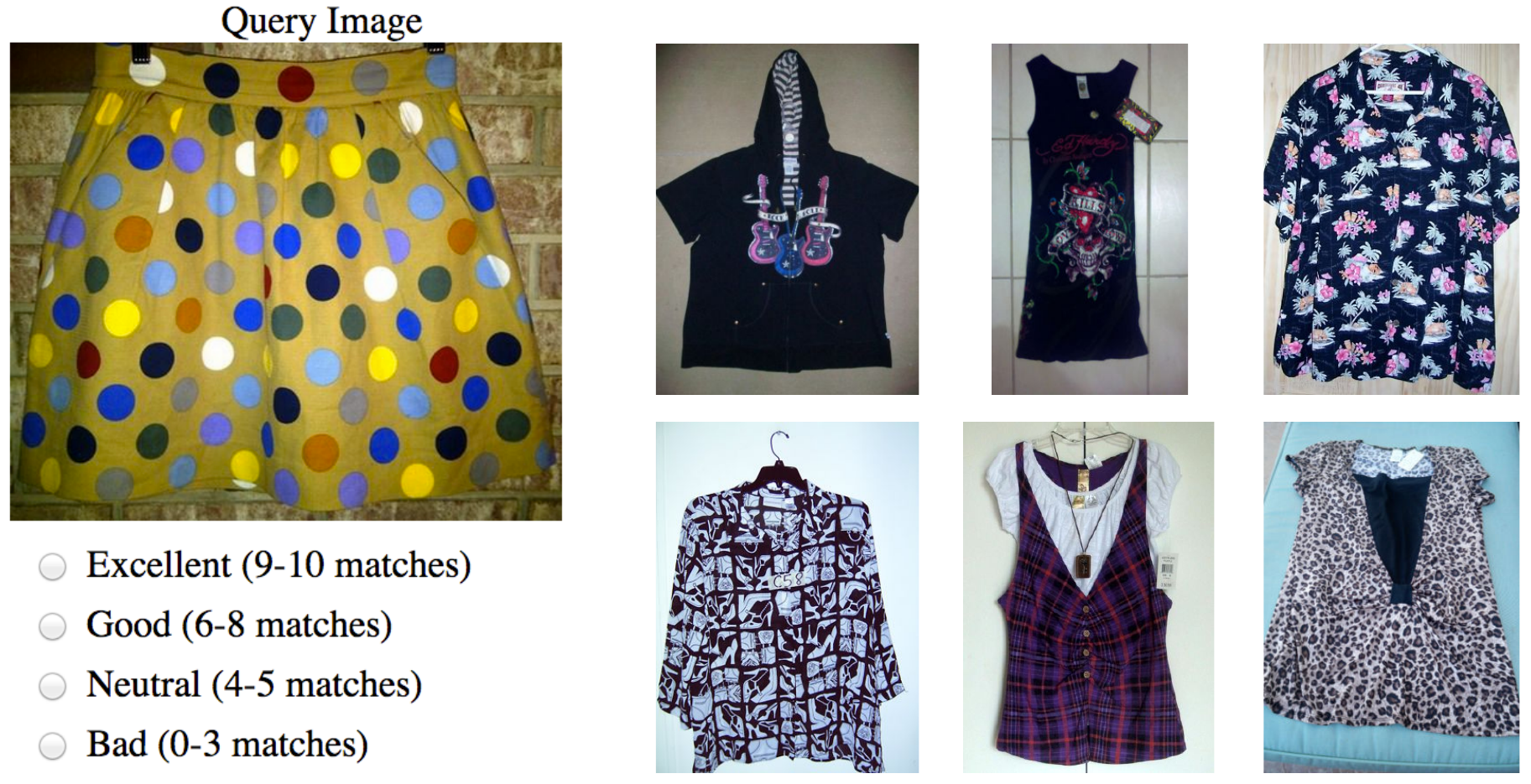} \\
\end{tabular}  
\caption{Sample Screenshot of Crowd-sourcing Experiment - A user is presented with a query image (polka dot skirt) and is asked to rate top-10 recommended tops on a 4-point scale (Excellent, Good, Neutral, Bad). For conciseness, only 6 results are shown in the figure.}  \label{fig:amtExperiments}
\vspace{-2ex}
\end{figure}
Few approaches have studied this problem from a vision science perspective~\cite{YamaguchiCVPR12}. These approaches propose a purely computational model of visual representation of fashion that completely ignores the role of users in the system. We believe this is a major limitation of these works as fashion experiences being social,cultural and visual can benefit greatly from users. Utilizing this information can often lead to better recommendations and improved user experiences. However, extending computational models to incorporate user feedback is non-trivial since understanding user choices in fashion is a challenging task. We propose to address these issues by taking a data-driven approach. Specifically, we answer following key questions in our study that leads us to very useful insights into solving this problem:\\
(i) \textbf{What can we observe from user preferences in fashion?}: Fashion choices can sometimes appear arbitrary or subjective. Hence a key aspect of understanding it is to take a data-driven approach that can generate useful insights averaged over a large population of users. Specifically, we take a closer look at large number of user preferences on visual recommendations. Not only does this framework allow us to quantitatively model key fashion concepts, it also enables us to qualitatively study them on real-world user feedbacks.\\
(ii) \textbf{How do these observations relate to vision science?}: A deeper dive into fashion understanding also requires us to study different domains or classes of fashion. Given our primary goal is to incorporate user preferences, a natural next step is to study if there are certain classes of fashion where these observations are more significant than others? If so, how can we automatically identify such classes? One interesting hypothesis that supports existence of these classes is rooted in the theory of visual perception. It postulates that human vision system works in layers of visual complexities~\cite{Serre05atheory} and an efficient model of machine perception can be built by taking these layers into account. \\
(iii) \textbf{Is it possible to build better computational models by integrating user insights?}: Once we have a reasonable understanding of how and why users prefer fashion, we are interested in integrating this information into our computational model. This could be done using State-Of-The-Art computer models inspired from techniques in Machine Perception. Our goal is to illustrate the robustness and value of this information for the task of visual fashion recommendation.In this paper, we present a systematic study of these fundamental questions and our key insights learnt through a large scale user study. The rest of the paper is organized as follows: Experimental Design section describes our proposed vision based computer models in detail and outlines datasets and framework for crowd-sourcing user preferences in fashion. Result section illustrates our findings from the experiments and also explains our recommendations in detail. Finally, we outline our conclusions in the last section.

\section{Experimental Design}
\label{sec:models}
In our current study, we have constrained the co-ordination choices to only two pieces in the outfits, namely skirts and tops. This is done purely for the sake of simplicity and the proposed techniques can be easily extended to co-ordinate more than two pieces in the outfit as well. Further, one of our primary objectives is to apply concepts from computer vision to model this task, hence we only exploit rich image data for modeling purposes. Using these constraints, we propose two models, DFR and SFR described below.

\subsection{Deterministic Fashion Recommender (DFR)}
Deterministic Fashion Recommender (DFR) aims to harness the power of data in recommending co-ordinating fashion items. For each co-ordinating clothing piece in our training data (e.g. \textit{skirts},\textit{tops}), we extract color features in form of a $K$-dimensional HSV histogram~\cite{Bhardwaj:2013:KDD}. This generates $40$-dimensional feature vector for each clothing piece. Features from $<$\textit{skirt},\textit{top}$>$ tuple are then indexed in a database. During testing time, when a query \textit{skirt} image is presented, a $40$-dimensional query feature vector is computed and searched across the \textit{skirt} portion of all indexed tuples. Finally, \textit{top} portions of all nearest tuples are returned as the recommended \textit{tops}. Since, the proposed model is based on a nearest-neighbor principle, the quality of recommendation depends on the size and quality of the data. With large and diverse data, DFR can provide better recommendations and can also be interpreted as being more objective with its fashion choices.

\begin{figure}[t]
\centering
\begin{tabular}{cc}
\includegraphics[width=.47\linewidth]{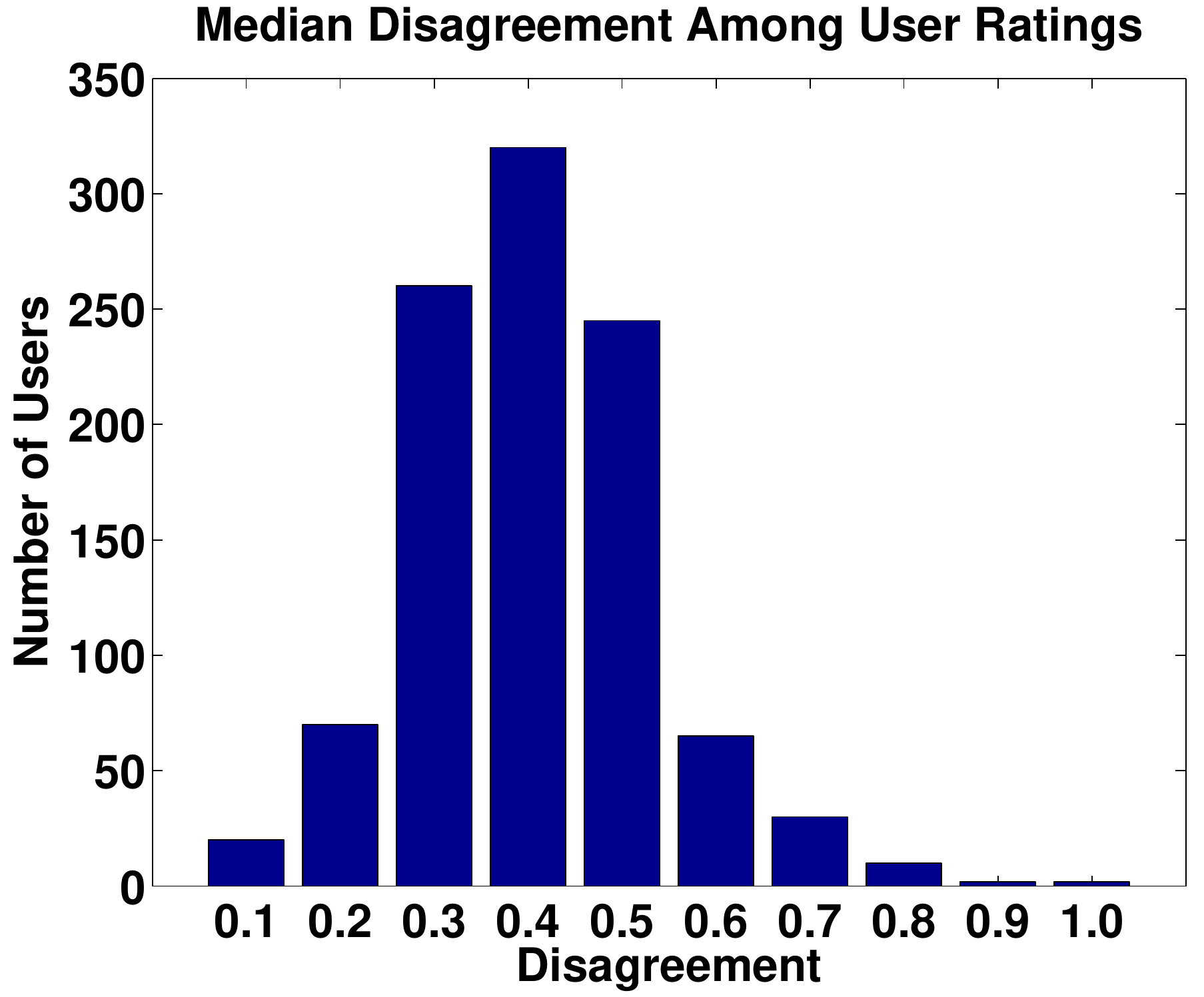}  &
\includegraphics[width=.47\linewidth]{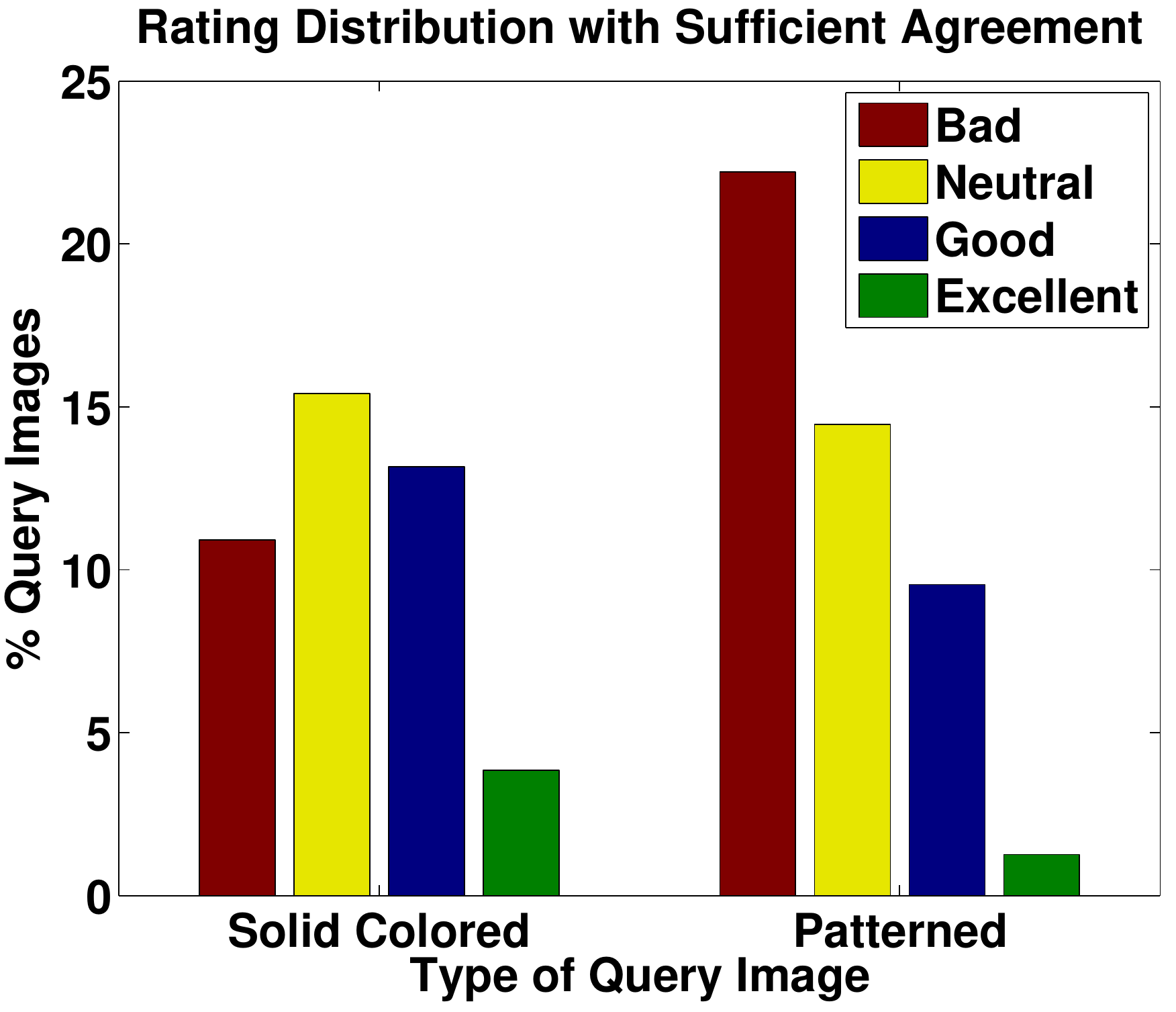} \\
(a) & (b)
\end{tabular}  
\caption{(a) Median Disagreements among user ratings. Majority users have agreeable ratings on fashion choices.(b) Depicts common agreement amongst users based on retrieved results. Results for solid colored queries are generally more favorable than for patterned queries.}  \label{fig:objExperiments}
\vspace{-2ex}
\end{figure}

\subsection{Stochastic Fashion Recommender (SFR)} 
Fashion choices can sometimes be highly subjective. For instance, a typical user may like polka dots so he or she may always prefer a match with polka dots over other matches. Such behavior cannot be modeled with DFR-like approaches. To address this issue, we propose Stochastic Fashion Recommender (SFR) that aims to explore the space of user biases in recommending matches. Computationally, the space of all such possible biases is huge and impractical to model. Hence, we constrain these biases to a smaller set of fashion choices. In our study, this constraint is achieved by the common notion that \textit{solid} and \textit{patterned} clothing co-ordinate well together. In other words, having busy patterns in both top and bottom clothing is less popular. To model this, we parameterize our desired output space (e.g. \textit{solids}) with a probability distribution. This ensures that given a \textit{patterned} query, all output choices from the model will be \textit{solids}. Given this particular output space, user biases can be modeled by a distribution sampling process. In our case, we perform a uniform sampling of the distribution (i.e. randomly pick a solid color). Hence, not only does this model ensure ``stochasticity" for the same input, it also relaxes computational complexity and allows us to closely mimic the human subjectivity in fashion.

\begin{figure*}[t!]
\centering
\begin{tabular}{ccc}
\includegraphics[width=.32\linewidth]{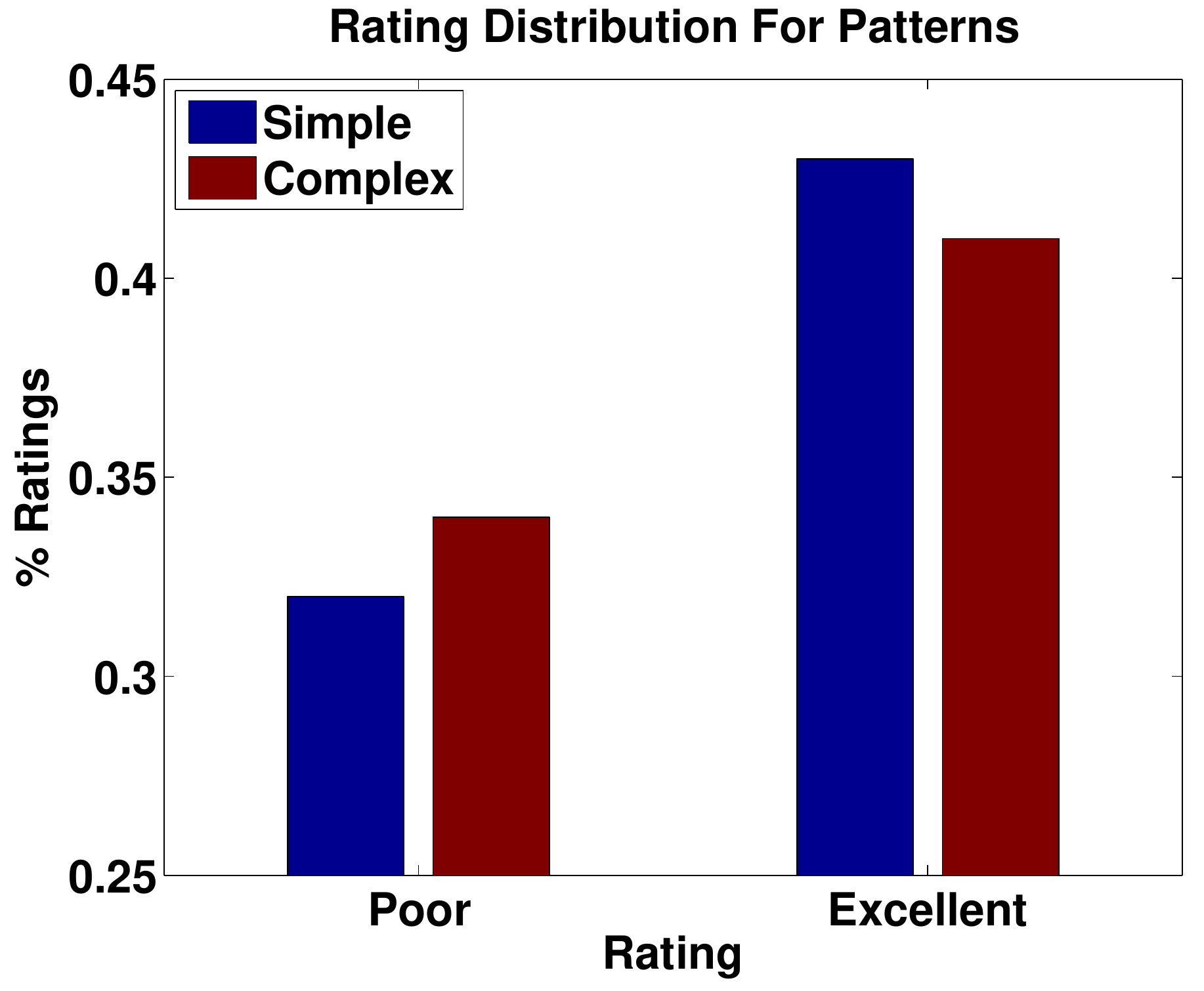} &
\includegraphics[width=.32\linewidth]{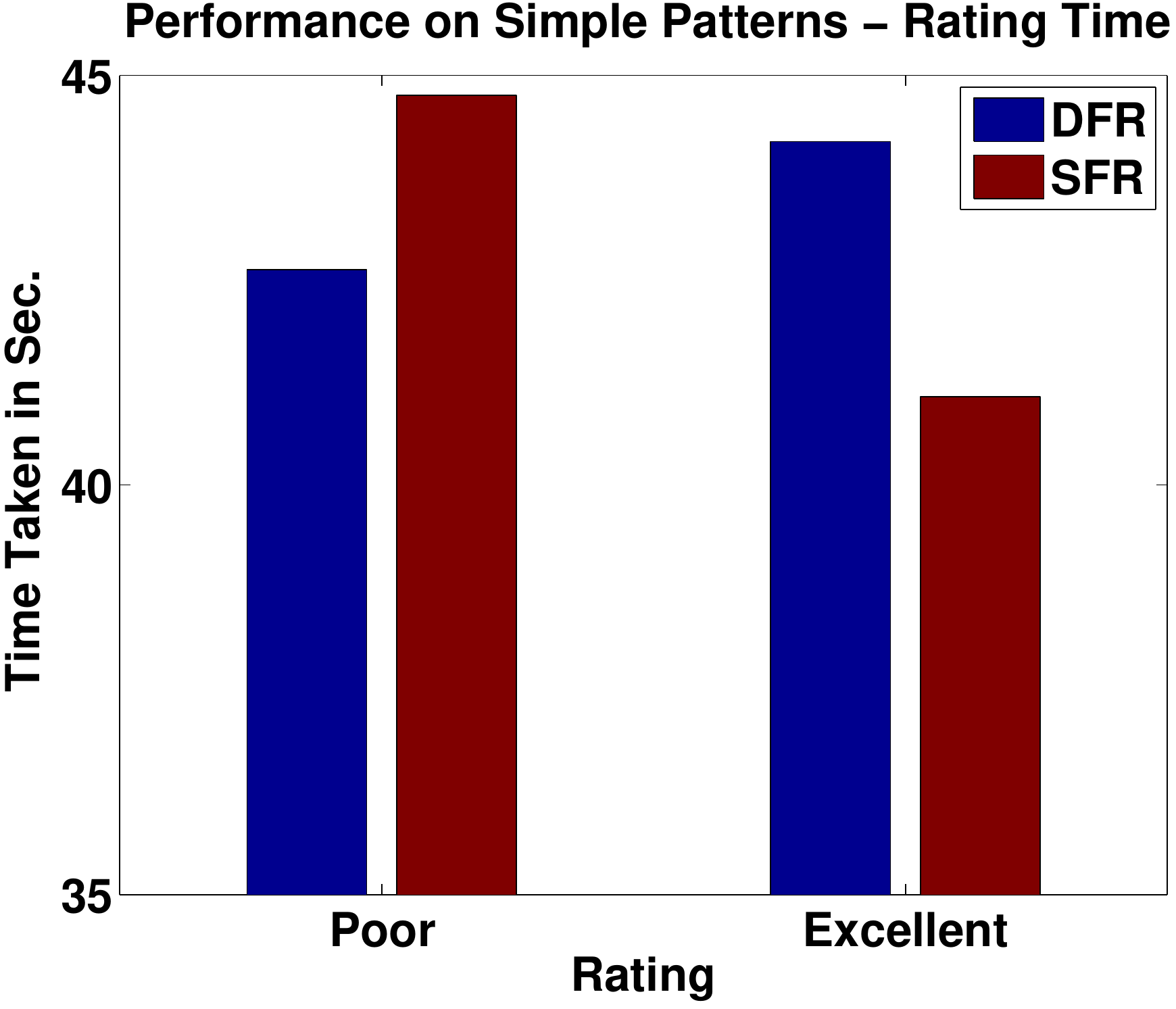} &
\includegraphics[width=.32\linewidth]{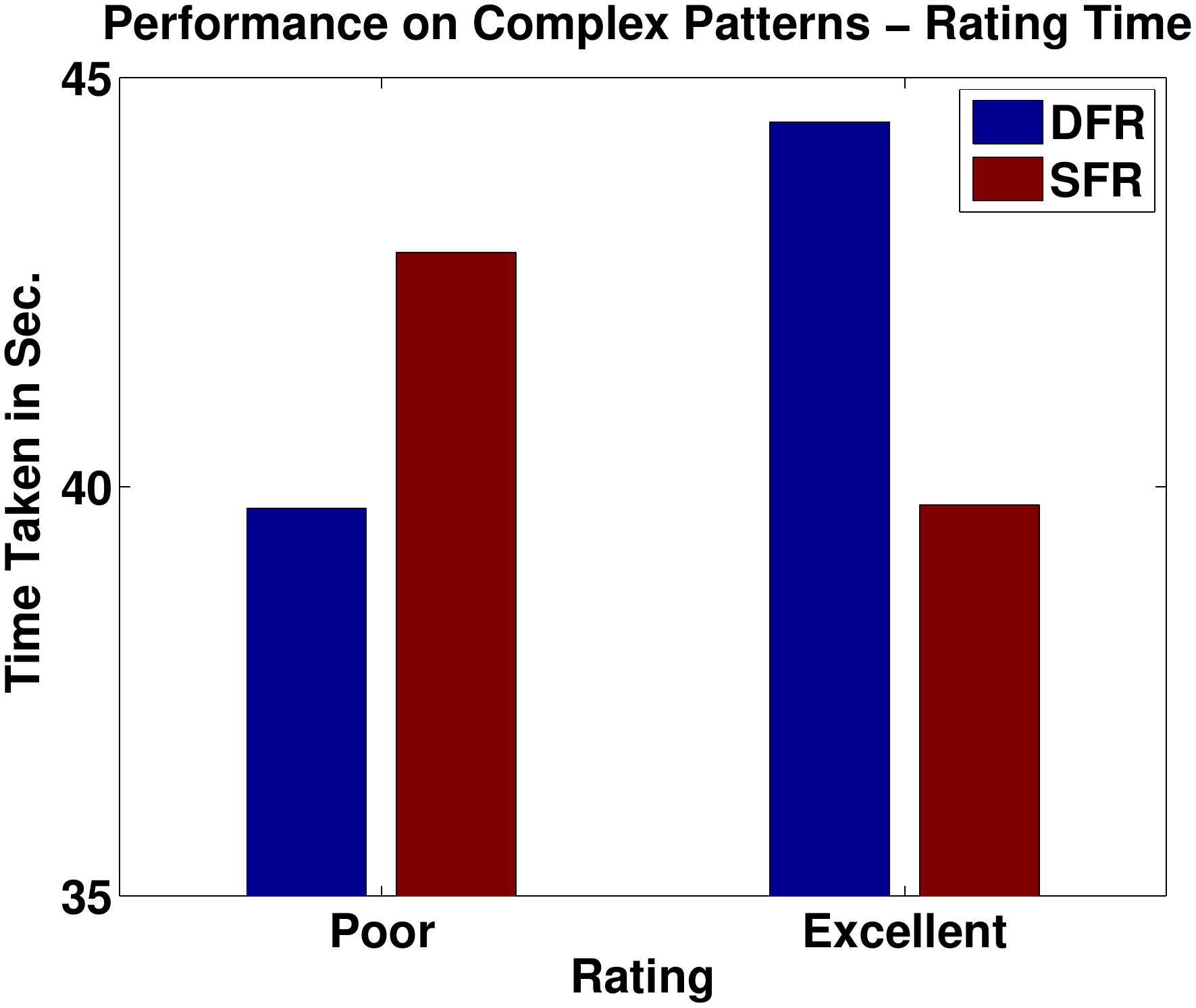} \\
(a) & (b) & (c)
\end{tabular}  
\caption{(a) Rating Distribution. Simple patterns are rated higher than complex patterns.(b) Model Performance on Simple Patterns. Matches from SFR are visually more appealing and hence take much less time for rating as compared to DFR. (c) Model Performance on Complex Patterns. SFR matches for complex patterns are more appealing as compared to simple patterns. }  
\label{fig:modelTimeExperiments}
\vspace{-2ex}
\end{figure*}
\subsection{Datasets}

We use $3$ datasets in our experiments, namely: \textit{Fashion-136K}, \textit{Fashion-350K} and \textit{Fashion-Q1K}. \textit{Fashion-136K} is a data set created by crawling the web for photographs of fashion models. It consists of 135893 street fashion images with annotations by fashionistas, brand, demographics. Hence, all images comprise top and bottom clothing co-occurring in the same image. This dataset is used for training both DFR and SFR models as described previously. \textit{Fashion-Q1K} dataset consists of $1000$ skirt images which are used as input queries to the models. Given these queries, both DFR and SFR models search and return recommended matches from \textit{Fashion-350K}. The result of retrieval is a ranked list of the Fashion-350K images sorted by relevance to a query from Fashion-Q1K. \textit{Fashion-350K} images are from a clothing inventory containing only top clothing (without model or mannequin).
\subsection{Crowd-sourcing Fashion}
Since the primary objective is to understand user preferences in Fashion, we require user ratings on recommendations provided by computer models DFR and SFR. We propose to use crowd-sourcing framework to obtain these ratings. This scheme also aligns well with our goal of understanding ``Street Fashion", where a layman user's (i.e. turk on Amazon Mechanical Turk(AMT)~\cite{AMT}) rating can be used as a ground-truth dataset. Figure~\ref{fig:amtExperiments} outlines the proposed crowd-sourcing experiment. A user is presented with an input query image on left side (polka skirt) and the top-$10$ recommended matches (tops) on the right side. User rates these matches as one of -1 (bad match), 0 (neutral match), 1 (good match), 2 (excellent match).
This validation procedure is in line with the established validation protocols for image retrieval~\cite{yao2011classifying}. To ensure consistency in ratings, each query was evaluated by $5$ different users leading to a total of $5000$ users. A total of $140$ unique users participated in the experiment of which majority of users had a $100\%$ approval ratings from their previous AMT experiments.

\section{Results}

\subsection{Understanding Users}
Understanding users is a logical first step in designing a user-driven fashion recommendation system. This can be done by gathering useful and interesting signals from user preferences that can be captured to replicate user choices in Fashion. We study this effect by introducing the notion of disagreement between user preferences over visual recommendations. Our hypothesis is that presence of strong signals in user preferences will lead to lower disagreement among their preferences. Denoting each rating of $2$ algorithms (DFR, SFR) for a query $q$ using a $2$-dimensional vector $\underline{\zeta}_{qi} \in \{-1, 0, 1, 2\}^2$, where each entry is a rating from -1 to +2, the score for disagreement across users is defined as, 
$\gamma_{qi} = \sum_{j=1}^2 ||\underline{\zeta}_{qi} - \underline{\zeta}_{qj}||_{1}$. 
Further, an agreement threshold is defined to be the median of disagreement scores across all queries, $A_T = median(\gamma_{qi}); q \in [1, 1000], i \in [1, 2]$. User ratings are retained only if $\gamma_{qi} < A_T$. This is done to filter outlier ratings from observations to account for spam in crowd-sourcing experiments. Finally, a query rating confidence is computed as $C_q = \frac{\text{Number of Users Retained for a Query}}{\text{Total Number of
  Users for a Query}}$. The final rating $R_m$ for a model $m$ is computed as:
\begin{equation}
R_m = \sum_{q=1}^{1000} C_q \frac{ \sum_{i=1}^5 \zeta_{qi} (m) \delta(
  \gamma_{qi} < A_T) }{\sum_{i=1}^5 \delta( \gamma_{qi} < A_T)} \end{equation} where $\delta(x)$ is the Dirac-delta function, $\zeta_{qi} (m)$ refers to the rating provided by fashionista $i$ on query $q$ for model $m$. 
As shown in Figure~\ref{fig:objExperiments}(a), majority of the users have lower disagreement score (below $0.5$) suggesting that there is enough consensus among their fashion choices that can be investigated in more detail. We now take a closer look at all these ratings where users have sufficient agreement as determined empirically through a threshold over disagreement score. These ratings are further split across solid and patterned queries to further isolate the particular section of queries which have greater rating agreements. Results shown in figure~\ref{fig:objExperiments}(b) suggest that users agree more on patterned queries, evidenced by the higher magnitude of red bars on the good agreement bin. In the case of solids, the users agreed that the retrieval was either \emph{Neutral} or \emph{Good} (evidenced by the higher weights to the yellow and blue bars). However, on patterned queries, users overwhelmingly agreed that the retrievals were not favorable (evidenced by the higher weights on the red bar). This result has two major conclusions: (i) \textit{There is a strong consensus among seemingly arbitrary fashion choices made by users} (ii) \textit{It is possible to identify a specific category of visual recommendations (i.e. solid colored bottom and patterned top combination) where such consensus is stronger.}

\subsection{Insights From Visual Recommendations}
The previous results provide an interesting cue that deeper insights can be revealed by segmenting visual recommendations into fashion classes such as solids or patterned queries. Our goal is to further analyze this behavior to ascertain its possible causes. One possible explanation can be found in similar results studied in computational neuroscience and vision science. These results directly correlate the complexity of an image to the visual processing time as well as image aesthetics. We hypothesize that since images in different fashion classes have varying level of visual complexity, it should also reflect in the choices made by the users. To illustrate this effect, we generate a weak categorization of all $8$ patterns into $2$ categories - Simple Patterns (Polka, Solids, Stripes, Plaids) and Complex Patterns (Animal, Floral, Geometric, Paisley) inspired by simple and complex cells in visual cortex. Using this scheme, we further analyze user ratings on these two categories. Figure~\ref{fig:modelTimeExperiments}(a) demonstrates that a larger proportion of users provide excellent ratings for matches corresponding to simple patterns. We further use this categorization to baseline visual recommendations on both fashion classes. This result has the following conclusion: \textit{ Fashion preferences tend have a strong correlation to associated complexities in visual perception.}

\subsection{Enhancing Recommendations With User Understanding}
\begin{figure}[t!]
\centering
\begin{tabular}{cc}
\includegraphics[width=.47\linewidth]{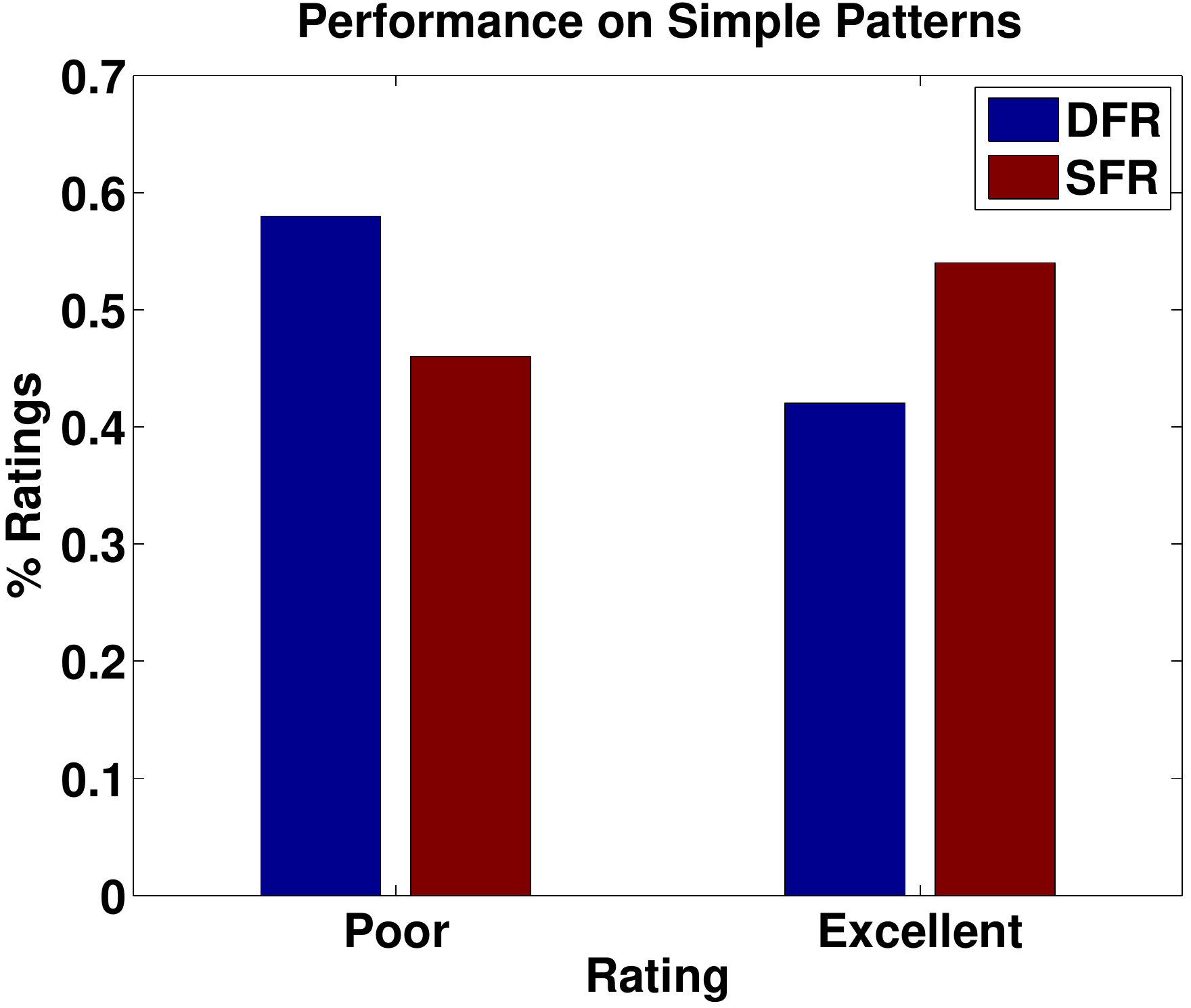} &
\includegraphics[width=.47\linewidth]{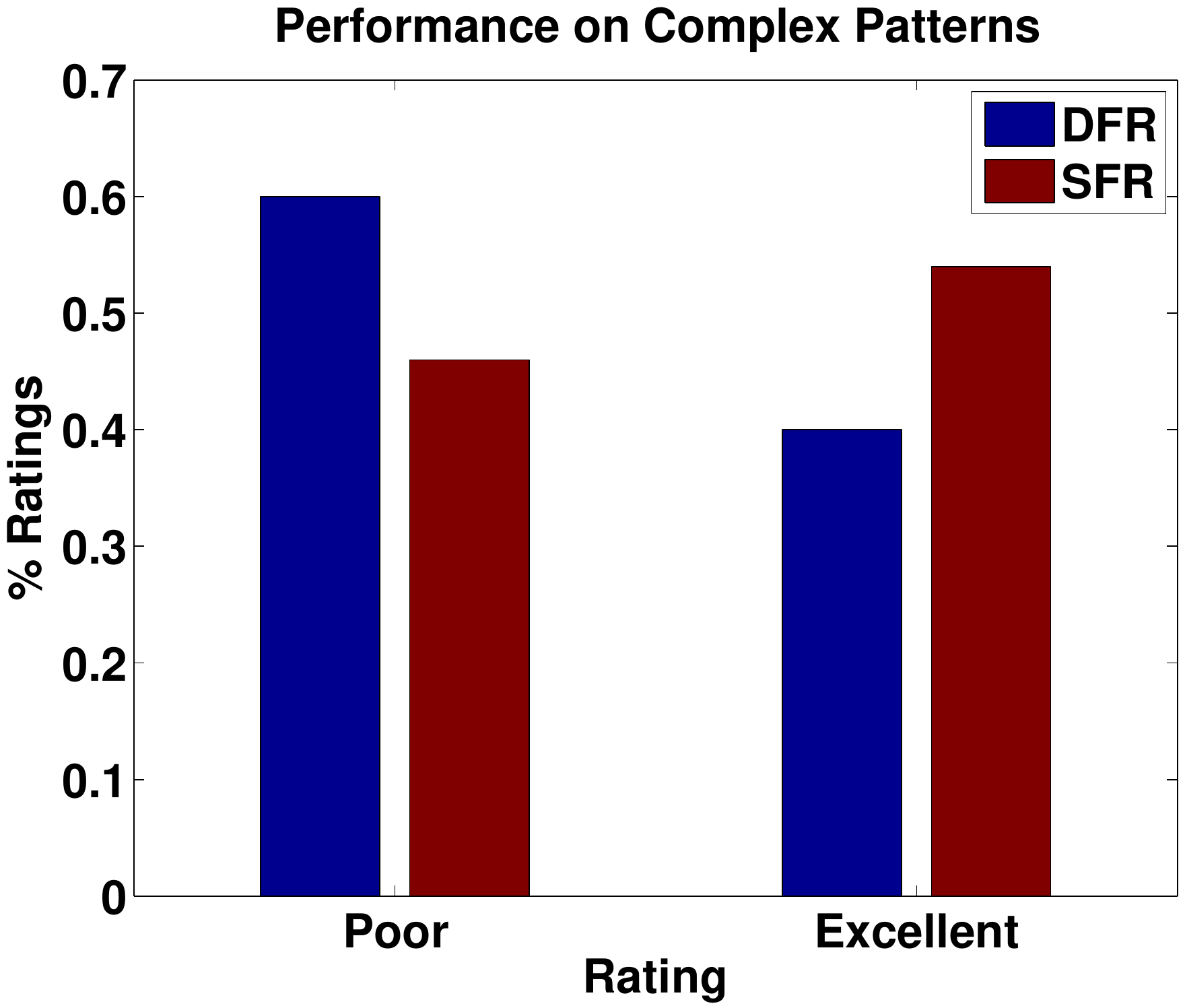} \\
(a) & (b)
\end{tabular}  
\caption{(a) Model Performance on Simple Patterns. Stochastic Models outperform (b) Complex Patterns.}  
\label{fig:modelExperiments}
\vspace{-2ex}
\end{figure}

All the above results provide a great insight into the role of users in fashion recommendation. However, they still do not answer if computer models can be built to replicate this understanding. To demonstrate this aspect, we evaluate two proposed models, DFR and SFR over the simple and complex pattern classes. Specifically, we look at two primary aspects, \textit{time to rate recommendations} and \textit{rating of recommendation} to illustrate this result. As shown in figure~\ref{fig:modelTimeExperiments}(b\&c), for all excellent ratings, users take much less time to rate recommendations from SFR as opposed to DFR. This is a strong indicator of the fact that visual recommendations from SFR have lower visual complexity and hence faster processing time for users. Moreover, the gain in time between SFR and DFR is even more pronounced for complex pattern classes. This again strengthens the argument that in the case of complex classes (when the visual complexity of the images are higher
), more elastic modeling techniques such as SFR provide better aesthetically pleasing visual recommendations as opposed to pure data-driven techniques like DFR. Hence, this result illustrates that users prefer visual recommendations from SFR over DFR models. Next, we look at the distribution of user ratings for both these models over different fashion classes. Figure~\ref{fig:modelExperiments}(a),(b) illustrate the performance of both the models. As shown, SFR outperforms DFR as it gets a much higher ratio of excellent user ratings and lower ratio of poor ratings for all the pattern classes. This result has following conclusions: (i)\textit{Incorporating user preference by incorporating stochasticity in recommendation model makes the results visually more pleasing and relevant to users.}(ii) \textit{Stochastic models are able to generalize well to different fashion segments as recommendations from both simple and complex patterns improve from it.}

\section{Conclusions}
In this paper, we presented a user-centric visual fashion recommendation system. The key results derived from this study suggest that users preferences often contain important behavior signals which correlate with associated complexities in visual perception and hence are useful for the task of visual recommendations. We show that by incorporating these signals into the design of a fashion recommender system substantially improves the quality of recommendations as well as provides visually appealing matches. Our future work focuses on deeper understanding of other aspects of fashion such as clothing style, finer-level texture mapping etc.

\bibliographystyle{acm-sigchi}
\bibliography{references}
\end{document}